\documentclass[
  journal=pasa,
  manuscript=research-paper,
  year=2025,
  volume=37,
]{cup-journal}
\usepackage{caption}
\usepackage{subcaption}
\usepackage{amsmath}
\usepackage[nopatch]{microtype}
\usepackage{booktabs}
\title{Streams and Shells Decoded: A Density-Driven Approach to Stellar Clustering in Galactic Halos with AstroLink}

\author{Viraj Ekanayaka}
\affiliation{Joint First Author}
\alsoaffiliation{Sydney Institute for Astronomy, School of Physics A28, The University of Sydney, NSW 2006, Australia}
\author{Smrithi Gireesh Babu}
\affiliation{Joint First Author}
\alsoaffiliation{Sydney Institute for Astronomy, School of Physics A28, The University of Sydney, NSW 2006, Australia}
\email[Viraj Ekanayaka]{naduran.ekanayakagedon@sydney.edu.au}

\author{William H. Oliver}
\affiliation{Interdisziplin\"{a}res Zentrum f\"{u}r Wissenschaftliches Rechnen, Universit\"{a}t Heidelberg, Im Neuenheimer Feld 205, D-69120 Heidelberg, Germany}
\alsoaffiliation{Zentrum f\"{u}r Astronomie, Institut f\"{u}r Theoretische Astrophysik, Universit\"{a}t Heidelberg, Albert-Ueberle-Stra{\ss}e 2, D-69120 Heidelberg, Germany}

\author{Geraint F. Lewis}
\affiliation{Sydney Institute for Astronomy, School of Physics A28, The University of Sydney, NSW 2006, Australia}

\newcommand{\block}[1]{}
\begin{document}

\begin{abstract}
We present a novel method to differentiate stream-like and shell-like tidal remnants of stellar systems in galactic halos using the density-based approach of the clustering algorithm AstroLink. While previous studies lean on observation, phase-space, and action-space based criteria for stream and shell determination, we introduce AstroLink's ordered-density plot and cluster identification as a viable tool for classification.
For a given data set, the AstroLink ordered-density plot reveals the density-based hierarchical clustering structure from which the resultant clusters are identified as being statistically significant overdensities.
Using simulations of sub-halo disruptions in an external potential to generate samples of tidal structures, we find that the curvature of the ordered-density plot is positive for stream-like structures and negative for shell-like structures. Comparisons with more standard classification techniques reveal strong agreement on which structures typically fit into stream-like and shell-like categories. Furthermore, we investigate the properties of clustered stream and shell samples in radial phase space and energy-angle space. Given the sensitivity of stellar tidal structures to their host dark matter halos, the identification and subsequent classification of these structures provide exciting avenues of investigation in galactic evolution dynamics and dark matter structure formation.
\end{abstract}

\section{Introduction}
Galaxies are surrounded by a dynamic array of smaller stellar structures, each following often complex orbits. Our very own Milky Way serves as an example with the Large Magellanic Cloud \citep[LMC;][]{Alves_2004}, globular clusters \citep{Harris_2010} and other dwarf satellites \citep{Battaglia_2013}, occupying the surrounding space. The life of these smaller stellar structures can be fleeting with the gravitational forces of the host galaxy tearing apart these objects, leaving behind tidal remnants that eventually fade into the background density. This process of accretion underpins the currently prevailing posit that galaxies form and grow hierarchically \citep{Subramanian_2000,Puebla_2024}. Naturally, such tidal remnants contain a wealth of information about the host galaxy and its history of evolution.

These remnant structures are primarily classed into stellar streams and shells. Streams are typically characterised by leading and trailing arms attached to the remaining core of the disturbed satellite. Though, more disrupted streams appear as a line of particles with no obvious core remnant. Shells appear as concentric circular arcs with a sharp density drop off at a certain radius. Intermediate structures that may also appear are sometimes placed into more niche classifications including umbrellas, tails and other structures \citep{Delgado_2010,Hendel_2015,Muller_2019,Carretero_2024}. The sensitivity of tidal remnants to the potentials of their host galaxies makes them an excellent probe of dark matter halos \citep{Bovy_2016,Bonaca_2018,Errani_2015,Law_2010}.

The inherent dependence on eye classification of these Low Surface Brightness (LSB) structures emphasise the need for automated codes for detection and classification of structures \citep{Atlas3d, Tal_2009}. In this paper, we introduce a tidal structure classification method using AstroLink \citep{astrolink}. With it, we demonstrate the ability to divide structures into two major categories, stream-like and shell-like structures. This paper serves as a proof of concept, intended to support and guide subsequent work.

The structure of this paper is as follows: In Section \ref{sec:background}, we provide an overview and background on stellar streams and shells. Section \ref{sec:method} describes the simulation–AstroLink–analysis pipeline. The classification method and clustering results are presented in Section \ref{sec:results}, along with validation of the identified structures in both radial and action space. In Section \ref{sec:discussion}, we discuss our classification with results from the literature and present our conclusions, outlining potential directions for future work.

\section{Background}
\label{sec:background}

Dark matter (DM), a mysterious substance that dominates the gravitational action of the Universe, has remained an intense topic of discussion since its initial discovery in the early twentieth century. Although there are anomalies and unanswered questions \citep{Peebles_2024}, the popular sentiment currently is that our Universe is described by the $\Lambda$CDM paradigm \citep{Blumenthal_1984}.  DM interacts predominantly through gravity, with negligible coupling to baryonic matter otherwise. Adding to this, its gravitational influence typically only becomes apparent at scales of order $\sim$kpc. As a result, it follows that, any baryonic structures that respond sensitively to the underlying gravitational potential serve as valuable tracers of their host dark matter halos. 

Tidal structures arise from the gravitational disruption of subhalos such as dwarf galaxies and globular clusters under the influence of the significantly deeper potential wells of their host galaxies. Streams, shells, umbrellas, plumes, tails are many of the named tidal structures that can be formed in various stages of the accretion process \citep{amina_streams, cooper_2010}. Tidal structures are predominant in early type galaxies (ETGs) where minor and intermediate mergers are prevalent \citep{Tal_2009}. Surveys suggest that around 10–20\% of nearby massive ellipticals show clear tidal debris \citep{Atkinson_2013}, although the true fraction may be even higher, hidden below detection limits. 

\subsection{Stellar Streams and Shells}
Stellar streams are long thin trails of stars and typically possess a small dense core, the non-disrupted remains of the original sub-halo and a leading and trailing stream of particles \citep{bullock_johnston, Amorisco_2015, Mancillas_2019, Valenzuela_2024}. More disrupted streams however, will not show an obvious arms and core remnant formation, and instead appear more as a single stream of particles. 

Another distinct archetype of tidal structures are stellar shells, interleaved arcs formed around the host potential. Shells are mostly aligned along the major axis of prolate galaxies while oblate galaxies with shells are azimuthally located near the equatorial plane \citep{Durpaz_1986, Hernquist_1987}. 
Typically, shells are categorised into three distinct subtypes, dependent on their spatial distribution properties \citep{Prieur_1990}. Namely;
Type I Shells: These are typically observed on opposite sides of a galaxy when viewed face-on, forming an aligned system. NGC 3923 in particular exhibits this particular composition \citep{2016A&A...588A..77B,1988ApJ...326..596P}.
Type II Shells: Also referred to as all-around systems, these shells form a nearly continuous envelope around the host galaxy. NGC 474 is a prime example of this configuration \citep{2020A&A...644A.164F, Bilek_2022}.
Type III Shells: The most ambiguous descriptor of the options, type III represents shells without clear alignment and thus often too vague to be confidently classified. This allocation is reserved for structures such as the interacting NGC 201-196 system.

Differentiating stellar streams and shells is important as it aids in understanding the merger history and properties of the progenitor. The presence of shells in galaxies would imply the progenitor was more massive and radial accretion while streams would follow a rather slow tangential accretion from less massive satellites. The morphologies of stream and shell like structures and the effect of the host halo and disk mass is studied in \citep{StreamGen}, where stream like structures were found to favour low energy orbits. Several instances of constraining dark matter masses of the host galaxy and the radial mass profile can be seen in the works of \citep{constraining_dm_streams,constraining_dm_satellites}.

The observation and categorisation of tidal structures in the local neighbourhood is a difficult affair. The galactic disk, in particular, proves to be a significant hindrance. Any streams or shells in the latter stages of accretion are homogenised enough to be virtually indistinguishable from the galactic disk. Adding to this, any structures near the disk are masked by the noise created from the stellar density of the disk \citep{Delgado_2010, virgo_diffuse_light}. As a result, observations are typically limited to tidal features that are still in the early to intermediate stages of disruption and located relatively at higher latitudes from the disk.

Despite these difficulties, the detection of these LSB structures has made remarkable progress in recent years, largely thanks to deep wide-field surveys like the Sloan Digital Sky Survey \citep[SDSS;][]{SDSS}, utilised photometric and spectroscopic telescopes to map a quarter of the sky close to the North Galactic Cap. The photometry involved the imaging of the sky under \textit{u, g, r, i} and \textit{z} filter bands \citep{Fukigita_1996} detecting wavelengths from $3000 \text{\AA}$ to $11000 \text{\AA}$. SDSS has been notably significant in detecting multiple substructures, including the Sagittarius \citep{saggitarius_stream,Belokurov_2006}, Orphan \citep{Grillmair_2006_orphan} and GD-1 \citep{Grillmair_2006_gd1} streams. Other relevant structures such as the tail of the Paloma 5 stream \citep{Odenkirchen_2001}, the Monoceros ring \citep{Newberg_2002}, and the Hercules-Aquila cloud \citep[HAC;][]{Belokurov_2007} were also identified from the data release. 
Following the success of the SDSS are other notable ventures, Dark Energy Survey \citep[DES;][]{DES}, Gaia \citep{GAIA} and Southern Stellar Stream Spectroscopic Survey \citep[$S^5$ Collaboration;][]{S5}. These have led to some notable findings; Orphan and Chenab streams are concluded as being products of a single progenitor \citep{Koposov_2019}, as have ATLAS and Aliqa Uma, making up the ATLAS-Aliqa Uma (AAU) stream. The disconnectedness of the latter has been attributed to the effects of the Sagittarius dwarf and Large Magellanic Cloud (LMC). A close interaction with Sagittarius produced a kink resulting in the single stream appearing as two. The effect was further exacerbated by interaction with the LMC which produced a misalignment in the stream track and proper motion giving the illusion of two separate entities. The HAC and VOD \citep{Virgo_overdensity} are two distinct structures that stemmed from the same accretion event \citep{Hercules_Aquila_virgo_clouds}.

Numerical simulations of shell formations suggest that these structures originate from minor merger events \citep{1990MNRAS.243..199H,formation_shells_hernquist} and provide valuable insights into the orbital dynamics of accreted stellar populations \citep{Hendel_2015}. The incidence of shell galaxies in n-body simulations of massive galaxies ($M_{200crit} > 6\times10^{12}M_\odot$) is estimated to be 
$18\pm 3\%$ \citep{Pop_2018}.

\section{Simulations and AstroLink}
\label{sec:method}
The study of tidal features has been significantly advanced through numerical simulations, with the development of clustering algorithms such as Subfind \citep{Subfind}, Velociraptor \citep{velociraptor}, Halo-OPTICS \citep{HALO_optics}, and CluSTAR-ND \citep{clustar_nd} serving as prime examples. These advancements are crucial in bridging the gap between theoretical models and real-world observations, particularly given the rarity of shell structures and the even greater challenge of observing them during their formation or evolutionary stages. \cite{classification_johnston} takes on an automated approach in characterising stellar halos trained and validated using a simulated dataset. It also addresses observational scenarios and possible shortcomings with observation. They employ the algorithm alongside the characterising system built using the morphology metric \citep{Hendel_2015}. Recent work by \cite{khalid_2024} introduced a classification system based on the visual characteristics of tidal structures and applied this framework to classify structures identified in various cosmological simulations. This study reaffirmed that the presence of these structures is not influenced by the subgrid physics of numerical simulations, indicating that they, as expected, result from gravitational interactions. The efficiency and timescale of this disruption and accretion process are largely determined by the initial orbital and structural properties of both the satellite and the host halo. Numerical simulations has greatly enhanced the ability of producing these structures and mimicking the observations conditions of the host and satellite which also aided in understanding of how galaxies grew overtime by mergers supporting the hierarchical model \citep{johnston_2008, Durpaz_1986, Subramanian_2000}.

\subsection{Sub-halo and Host Potentials}

Our analysis is conducted on a catalogue of tidal structures produced through the time evolution of synthetic sub-halos. Here, we present a sample population of 8 unique structures. Several factors motivate the utilisation of this idealised sample data set. There is intention here to mimic a possible tidal structure population around a galactic halo while preserving clarity in the structures formed. That is, the idealised population consist of completely disrupted sub-halos that form well defined structures. They are easily identifiable with visual inspection and form an important point of comparison. We have curated the initial conditions and evolution times to produce a range of streams and shells with slight differing characteristics to obtain a baseline of the structures' AstroLink ordered density distribution. Table \ref{IC_table} lists out the mass and evolution times for the initial sub-halos with further detail provided in Section \ref{sec:results}.

 \begin{table}[h]
\begin{threeparttable}
\caption{Summary of the initial conditions for evolved sub-halos. Each name corresponds to the final structure as seen in figure \ref{fig:four_scatter_plot}}
\label{IC_table}
\begin{tabular}{lccl}
\toprule
\headrow \textbf{Name} & \textbf{Mass} & \textbf{Evolution} & \textbf{Initial}\\
\headrow  & (${\rm M_\odot}$) & \textbf{Time (Gyr)} & \textbf{Orbit} \\ \cline{1-2}
\midrule
Stream 1 & $10^7$ & 4.2 & Elliptical \\ 
\midrule
Stream 2 & $10^7$ & 8.8 & Elliptical \\
\midrule
Stream 3 & $10^7$ & 10 & Elliptical \\
\midrule
Stream 4 & $10^7$ & 7.2 & Elliptical \\
\midrule
Stream 5 & $10^5$ & 4.9 & Circular \\
\midrule
Stream 6 & $10^5$ & 5 & Elliptical \\
\midrule
Shell 1 & $10^7$ & 5 & Radial \\
\midrule
Shell 2 & $10^7$ & 4.8 & Radial \\
\bottomrule
\end{tabular}
\block{
\begin{tablenotes}[hang]
\item[]Table note
\item[a]First note
\item[b]Another table note
\end{tablenotes}}
\end{threeparttable}
\end{table}

Ultimately, we aim to demonstrate a discernible difference in shell and stream like structures which may be utilised to form the basis of a new classification methodology. This will be further discussed in Sections \ref{subsec:astrolink} and \ref{sec:results}. The sub-halos orbit in a Milky Way-like potential and evolved for varied time lengths. For this purpose, we choose Plummer spheres for the sub-halo progenitors for which the potential is given by 
\begin{equation}
    \Phi_{\text{Plummer}} = -\frac{GM_{\text{P}}}{\sqrt{r^2 +R^2}},
\end{equation}
with total mass of the sub-halo $M_{\text{P}}$ and scaling parameter $R$ which characterises the radius at the density fall off. 

For numerical initialisation of the Plummer spheres, we follow the methodology of \cite{Aarseth_1974}. This provides unit spheres which we adjust for desired sub-halo masses. We consider the particles to be dark matter-like and equivalent to stellar particles as the main intention is to serve as tracers of position-velocity distribution in streams and shells. They fixed particle mass at $10$ ${\rm M_\odot}$. Total mass is thus adjusted with particle number. We employ progenitor masses of $\sim10^5$ ${\rm M_\odot}$ and $\sim10^7$ ${\rm M_\odot}$. The majority of structures are in the order of $10^7$ ${\rm M_\odot}$ with the less massive sub-halos utilised for comparison in formed structures. Phase space (3D position and 3D velocity) conditions are arbitrary and adjusted for structure formation. 

The produced sub-halos are initialised and evolved via parallelised N-body simulation code GADGET-4 \citep{2021MNRAS.506.2871S}. We adopt the configuration of collisionless particles due to the relative scale of our simulations. We model the host galaxy's potential as a static potential with a Milky Way-like spherical bulge-disk-halo framework. Following the method of \cite{genetic_approach_MC}, the three components are characterised as follows. The bulge is defined as a Hernquist potential \citep{1990ApJ...356..359H},
\begin{equation}
    \Phi_{\text{bulge}} = -\frac{GM_{\text{bulge}}}{r_{\text{bulge}} +r}.
\end{equation}
For our purpose, we set radii scale $r_{\text{bulge}}=0.7$ kpc and bulge mass $M_{\text{bulge}}=0.7\times10^{10}$ ${\rm M_\odot}$. 
The disk is comprised of the Miyamoto-Nagai potential \citep{1975PASJ...27..533M},
\begin{equation}
    \Phi_{\text{disk}}(R,z)=-\frac{GM_{\text{disk}}}{\left(R^2+\left(r_\text{disk}+\sqrt{\left(z^2+b^2\right)}\right)^2\right)^{1/2}},
\end{equation}
where $b$ is the scale height which we set to a fifth of the disk radius. Radii scale $r_{\text{disk}}=3.5$ kpc and the disk mass is set at $M_{\text{disk}}=0.7\times10^{10}$ ${\rm M_\odot}$. 

Finally, the encompassing DM halo is represented as the typical Navarro-Frenk-White (NFW) potential, 
\citep{navarro1997apj}, defined as:
\begin{equation}
\Phi_{\text{halo}} = -\frac{GM_{\text{halo}}}{r} \ln\left(\frac{r}{r_{\text{halo}}}+1\right).
\end{equation}
The spherical halo's virial radius is given by $r_{\text{halo}}=280$ kpc and mass $M_{\text{halo}}=1.3\times10^{12}$ ${\rm M_\odot}$. The mass and radii are derived from \citep{genetic_approach_MC}. The parameter values for the other components are taken from  \cite{Joss_2016} and \cite{MW_McMillan}.

In comparison to the composite model described in \citep{Bovy_Galpy}, both frameworks adopt a three-component bulge–disk –halo structure, with identical functional forms for the disk and halo potentials. The Galpy implementation, referred to as MWPotential2014 \citep{galpy}, employs a power law profile with the density profile exponentially cut off at exponent of 1.8 and a radius cut of at 1.9 kpc. In contrast, the model adopted in this work features a spherical bulge represented by a Hernquist profile, as previously outlined. A comparison of the mass and radial parameters of the bulge, disk, and halo components highlights that while the bulge masses are comparable, the disk is more massive in the Galpy model, whereas the halo mass is dominant in our presented model.

\subsection{AstroLink}
\label{subsec:astrolink}
In order to identify and analyse the structures present in our simulations, we employ AstroLink \citep{astrolink} -- the successor of Halo-OPTICS \citep{HALO_optics} and CluSTAR-ND \citep{clustar_nd}. AstroLink is a hierarchical density-based clustering algorithm that identifies structures via their statistical distinctiveness from noisy density fluctuations inherent within the data. The data can be of any size and dimensionality, and similarly AstroLink can find an arbitrary number of arbitrarily sized and shaped clusters -- with the only working assumption being that it is meaningful to assess the \textit{similarity} of neighbouring points with a Euclidean (or Mahalanobis) distance. Compared to its predecessors and to traditional halo-finders like AHF \citep{Gill_2004, Knollmann_2009}, AstroLink has been shown to be better at finding infalling and disrupted satellite remnants within simulated galaxies while also taking less run-time \citep{astrolink, Oliver2024_neurips}.

In short, the clustering process of AstroLink is as follows:
\begin{enumerate}
    \item If the hyperparameter \texttt{adaptive} $=1$ (default), AstroLink will first transform the data so that it has unit variance in each dimension. This removes the unwanted effect of having differing units within the data.
    \item The local-density of each point in the (possibly transformed) data is then computed using its $k_\text{den}$-nearest- neighbours ($k_\mathrm{den}=20$, default). The logarithm of this quantity is then taken and rescaled so that $\log\hat\rho \in [0, 1]$.
    \item The data points are then connected together via their $k_\text{link}$-nearest-neighbours in order of decreasing $\log\hat\rho$. By default $k_\text{link}$ is determined automatically using $k_\text{den}$ and the dimensionality of the data. This step forms a hierarchy of groups and the ordered-density plot -- a plot of $\log\hat\rho$ vs an ordered-index as seen in Figure \ref{fig:odplot}.
    \item A model is then fit to the set of group prominences -- a measure of their overdensity -- from which clusters are identified as those groups that are $S$-sigma outliers in this distribution. The default value of $S$ is calculated as the expected maximum sigma value of $n_\text{groups}$ samples drawn from the now-fitted model.
    \item Finally, if the hyperparameter \texttt{h\_style} $=1$ (default), a correction is applied to the hierarchy clusters by incorporating some additional outlier overdensities -- producing the final hierarchy of clusters.
\end{enumerate}
For more algorithmic details, we refer the reader to the original science paper \citep{astrolink}, as well as to the AstroLink GitHub\footnote{https://github.com/william-h-oliver/astrolink} page.

Although we do use AstroLink to identify tidal structures within this work, it is the AstroLink ordered-density that is the more novel contribution to our method. Any structure identifiable with AstroLink will appear as a contiguous peak within the distribution of ordered-density. The shape of this peak is directly related to the density profile of the structure within the feature space of the data (as estimated by step 2 above). Any structure whose density profile is centrally concentrated with heavy tails, or contrarily, flat with sharp boundaries, will appear as such in the ordered-density plot. As such, the shape of the region of the ordered-density plot that corresponds to a cluster contains information about its morphology. We apply this notion in Section \ref{sec:clustering_classification} to differentiate between stream-like and shell-like tidal structures.

\section{Results}
\label{sec:results}
\subsection{6D Clustering}
\label{sec:6d_clustering}
\begin{figure}[ht!]
\centering
\includegraphics[width=\linewidth]{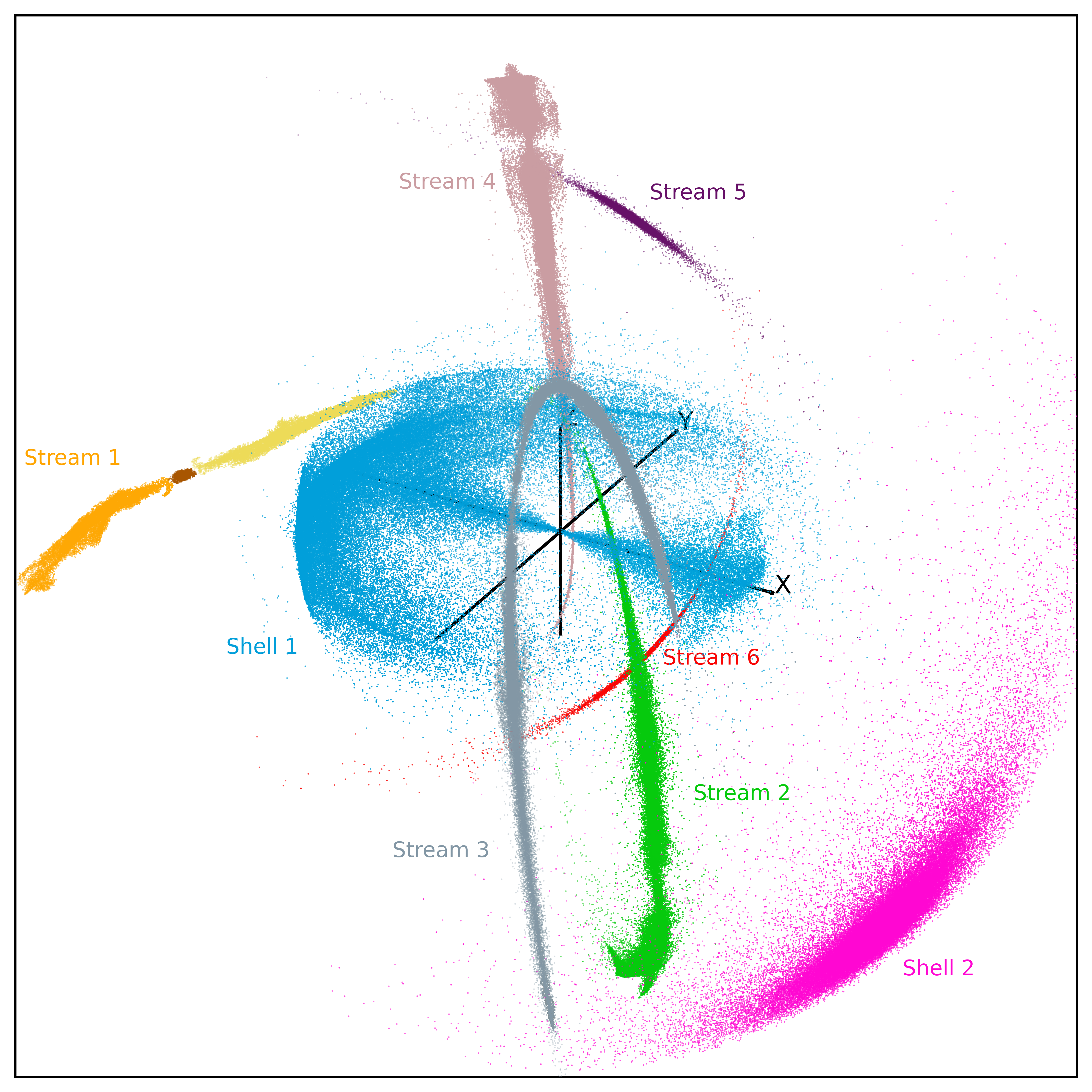}
\caption{Particle distribution of composite tidal structure population in 3D position space. Population is clustered with AstroLink in 6D (3D position and 3D velocity) space. Each colour represents an independent cluster. No sub-clusters are visualised. All clustering projections use AstroLink default parameters $k_\text{den}=20$ and $S=$`auto'. Each separate structure is visually classified here and is as labelled.}
\label{fig:four_scatter_plot}
\end{figure}

For the purpose of our analysis here, we present a composite tidal remnant population with distinct clusters identified via AstroLink. Figure \ref{fig:four_scatter_plot} visualises the population clustered in 6 dimensional (3D position and 3D velocity) space. All structures in the tidal structure catalogue are identified. The given labels are based on visual inspection of the obtained clusters. The tidal structures depicted are the result of sub-halos disrupted in a Milky Way-like potential centred at the origin and as detailed in Section \ref{sec:method}. For simplicity, we use the default AstroLink parameter settings to identify clusters which are expected to be near-optimal in most cases. Noting that the parameters $k_\text{den}$ and $S$ have the largest effect on the output, we have found that the results are robust to variations in the ranges of $[10, 30]$ and $[4, 6]$ respectively. While for smaller $k_\text{den}$ and/or $S$ the resulting clusters can be noise-dominated and for much larger $k_\text{den}$ and/or $S$ the clusters can fail to be detected.
\begin{figure*}[ht!]
\centering
\includegraphics[width=\linewidth]{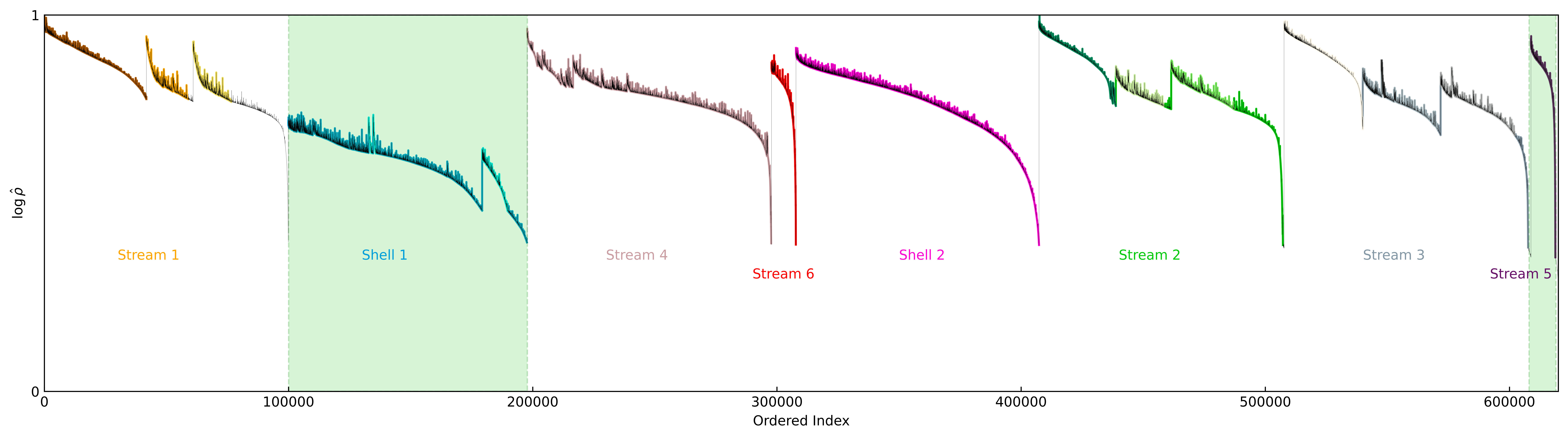}
\caption{Ordered density plot of the composite tidal population clustered in 6D (3D position and 3D velocity) space. Each colour as labelled corresponds to the cluster of the same colour presented in Figure \ref{fig:four_scatter_plot}. Highlighted green areas correspond to the structures presented in Figures \ref{fig:classify_stream}, \ref{fig:classify_shell}.}
\label{fig:odplot}
\end{figure*}
Since AstroLink finds a hierarchy of clusters, some streams are further segmented into sub-clusters as a result of them containing significant internal overdensities -- however no two disrupted subhaloes are found as a single cluster. In Figure \ref{fig:four_scatter_plot}, each colour represents a distinct independent cluster and all distinct structures are labelled. Here however, the lower order sub-clusters are not shown as they are considered sub-groups of the primary identified structures.

The central disk-like structure on the $x-y$ plane, `Shell 1', takes the form of a typical type 1 stellar shell system with formations on either side of the axis. With an initial mass of $10^{7}$ ${\rm M_\odot}$, the shell's sub-halo progenitor was evolved for $5$ Gyrs with an initial velocity of $200$ km/s directed $away$ from the origin (i.e. the centre of the host potential). We observe a separation between the primary shell and a `trailing' shell produced through the influence of the host's central bulge, coupled with the stronger influence of the halo and disk components near the centre. Shell 2 is present well outside and away from the busy regions of tidal system. Unlike the central shell system, Shell 2 is not accompanied by any other shell substructure, categorising it as a Type 2 shell system. The velocity is primarily directed along a line through the origin. However, unlike the central shell, the velocity of Shell 2 is directed $towards$ the origin with a bias in the $x$ direction. 

Ultimately, both structures were subjected to extreme eccentric orbits near the core of the host potential subjecting them to tidal shock. Consequently, the central shell in particular, adopts a `fanned out' particle distribution with no discernible remains of the progenitor core. As such, we end up observing the characteristic partial spherical shell-like distribution. Notably, the radii of the particle distribution is constrained with a sharp drop-off dictated by the initial conditions. As expected, we find that a greater radial distance and radial velocity leads to a greater maximum shell radius $r_{ms}$.

Besides the two shell structures, the remaining accompaniment of structures are thin curved arcs traditionally classed as stellar streams. With a closer inspection, we find that the graded orange-yellow structure, `Stream 1', is perhaps the most typical stream-like structure. It possesses leading and trailing arms attached to a central core that has yet to be completely disrupted. The three sections of the stream are considered distinct, with AstroLink identifying them as separate structures. The progenitor originated with a mass of $10^{7}$ ${\rm M_\odot}$ and a velocity of $100$ km/s aligned on a $45$ degree offset towards the origin from a radial path. At the present stage, it has evolved for $\sim4$ Gyrs. 

Streams 2 and 4 were $10^{7}M_\odot$ satellites evolved for $8.8$ and $7.2$ Gyrs respectively, with initial velocities normal to the host potential. Despite the longer time exposed to disruption, both structures retain their primary stream characteristics. Both structures display identifiable leading and trailing streams with a core region in the middle. However, unlike the previously discussed stream, from the perspective of AstroLink, Stream 4 is considered a single structure without any significant internal clumping. The stream also exhibits a degree of radial flattening of the arms near the central core akin to that of a shell. AstroLink is able to distinguish between the core and arms of Stream 2, with the arms being identified as sub-clusters of the whole of the stream. The graded grey, Stream 3 is structured similarly, with the arms considered as sub-clusters. Streams 5 and 6, originating from $10^{5}$ $M_\odot$ sub-halo progenitors undertake a more complete disruption with no clearly recognisable core remaining in the streams. Both structures have evolved for $\sim5$ Gyrs. 

The orbits of each of these streams have low eccentricity. None of them approach the central core of the potential and thus are not subjected to the same tidal shock experienced by the shell progenitors. As a consequence, the majority of the presented streams are able to preserve a core remnant of their initial sub-halos. Furthermore, the particle contents morph into thin trails along the orbit of the satellite as opposed to the dissipated partial arcs of stellar shells. We note that the $10^{7}$ ${\rm M_\odot}$ progenitor streams all possess a discernible core despite evolution times ranging from $4-10$ Gyrs. On the other hand, as expected, the smaller $10^{5}$ ${\rm M_\odot}$ progenitor red streams have their core completely disrupted.

A closer inspection may be conducted with the assistance of AstroLink's output. Figure \ref{fig:odplot} displays the ordered-density plot attained from applying AstroLink to the composite structure's six-dimensional phase-space. We note several remarkable features: First is the separation between the core and arms of the available streams. The ordered index spanning from $0-100,000$ represents Stream 1. The first curve characterises the remaining core fragment. The proceeding two curves identify the leading and trailing stream arms respectively. Despite the separate clusters, the three stream clusters are considered part of the same density region. A sharp $\log\hat\rho$ density drop off proceeding the trailing arm cluster indicates a separation of density regions between cluster groups.

The core and stream arm separation may be similarly observed for Streams 2 and 3. However, as previously discussed, the separation is of lower order and the arms of the streams are considered sub-clusters as opposed to independent structures. Although we observe a spike in the density indicating a similar arm separation for Stream 3, the density difference fails to reach the significance cutoff for either a separate cluster or sub-cluster. Streams 5 and 6 are homogenised and display no distinct arm separation from any remaining central core remnant.

The shell components of the tidal population exhibit smoother $\log\hat\rho$ distributions. Presented in the first highlighted region in figure \ref{fig:odplot}, we see Shell 1, which possesses a second density sub cluster arising from the previously discussed trailing shell segment. Shell 2, which contains no such trailing structure, produces a smooth curve with no secondary components.

Importantly, we note the dichotomy of the profile of the ordered-density plots between streams and shells. The shells exhibit a profile with negative curvature while the streams possess a positive curvature. It must be acknowledged however, that the separable core portions of the streams possess a similar profile to that of a shell. Taking the curvature difference into account, we can independently categorise `stream'-like and `shell'-like structures entirely from the AstroLink produced ordered-density plot. We will discuss this further in Section \ref{sec:clustering_classification}, with Figures \ref{fig:classify_stream}, \ref{fig:classify_shell} providing a visualised cue of the difference as well as a breakdown of the categorisation methodology. The two structures presented correspond to the highlighted ordered-density plot in Figure \ref{fig:odplot}.

\subsection{Radial Phase-Space}

\begin{figure}[hbt!]
\centering
\includegraphics[width=\linewidth]{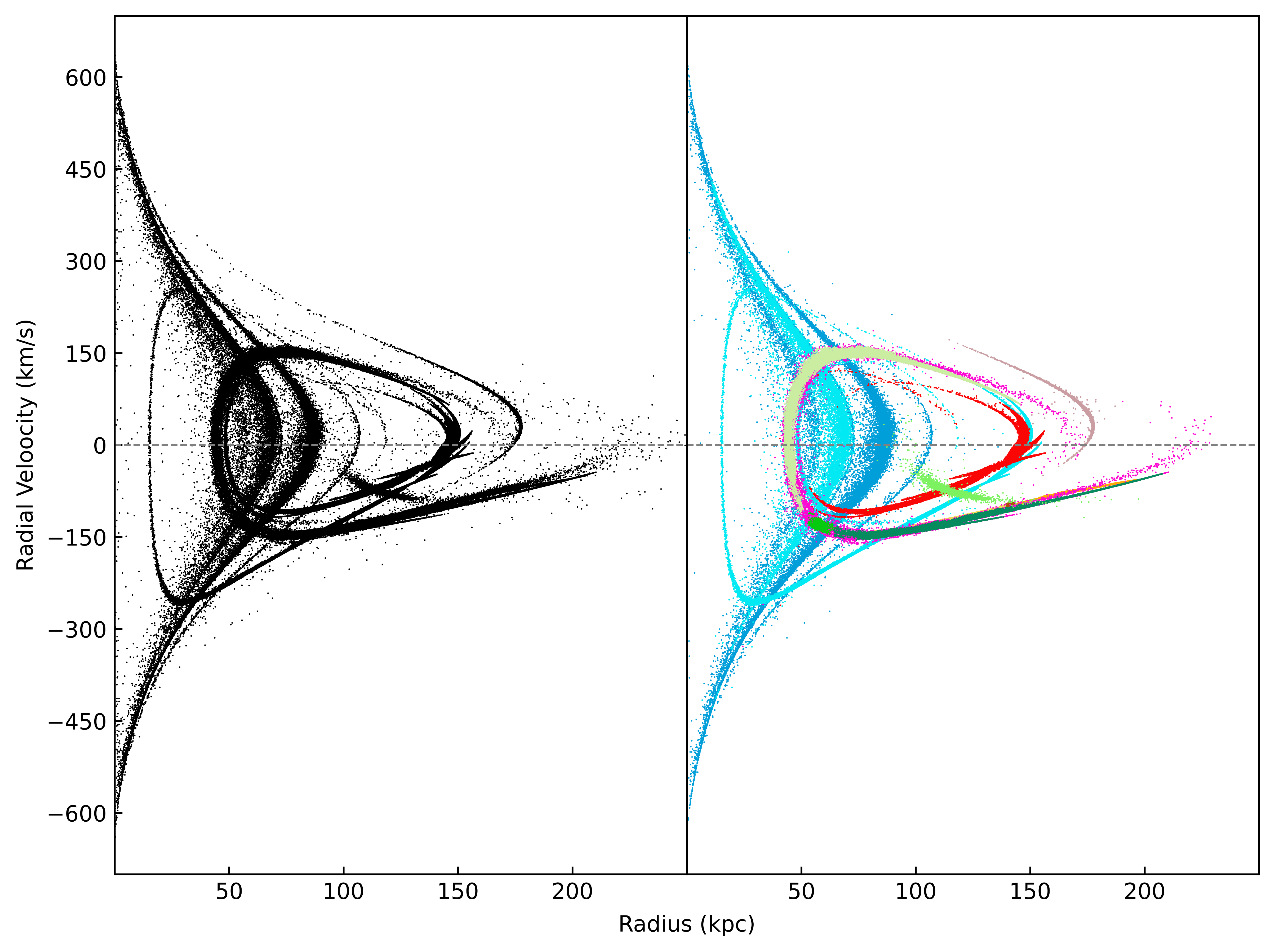}
\caption{Unclustered and 6D clustered radial phase distribution of the tidal composite population. All cluster colours match the corresponding structures in Figure \ref{fig:odplot}.}
\label{fig:radial_phase_space}
\end{figure}
Figure \ref{fig:radial_phase_space} depicts the unclustered and 6D clustered radial phase-space for the sample composite potential presented in Section \ref{sec:6d_clustering}. We see that the overlapping structure profiles of the unclustered phase space are resolved through the application of AstroLink. 

Clustered in the blue shades of the central Shell 1 and the accompanying sub-cluster shell, we see a radial velocity dispersion ranging between $\sim-600$ km/s to $\sim+600$ km/s, characteristic of a highly eccentric orbit with close proximity to the host's centre. That is, a greater acceleration near the host potential centre produces the large inward (negative) and outward (positive) velocity distribution. However, the majority of particles ($\sim70\%$ in the central shell) lie within a $100$ km/s range from $0$ radial velocity. This is attributed to the slowing down of the particle distribution at they reach the radial peak of the shell arc. We also observe the formation of multiple `strips' in the distribution. This arises from the presence of the trailing shell segment and formation of multiple sub arcs in the overall shell. 

Contrastingly, depicted by the streak of pink particles seen ranging past a $200$ kpc radius, Shell 2 exhibits a more concentrated profile. It is worth pointing out that in spite of the disparity in occupation area in radial phase-space, both shell structures originate from equal mass progenitors and the clusters contain similar particle counts. We suppose that this is due to the difference in orbits between the two structures. Shell 2, although possessing an eccentric orbit, does not cross the origin point as Shell 1 does. Consequently, the progenitor sub-halo does not experience the same tidal shock responsible for the `fanned out' particle spray experienced by the central shell. The result is a longer disruption time for the outer shell. This leads to a more consistent morphing of the sub-halo into the visualised shell. In summary, despite both possessing highly eccentric orbits, the shells form in two separate ways; 1. spraying out from tidal shock by passing through the centre, 2. a slower expansion from close orbit to the centre.

Turning our attention to the stream-like structures, we find that their radial phase-space profiles are more consistent, exhibiting a `guitar pick'-like shape. Notably, the radial velocity is more constrained with all but one stream contained within $200$ km/s from zero. The stream lying outside this boundary corresponds to Stream 4 previously discussed in Section \ref{sec:6d_clustering}. Although the clustered sample of Stream 4 (rose-brown particles) is seemingly constrained, an inspection of the corresponding unclustered profile shows an extension of the stream exceeding $300$ km/s. Interestingly, it is the most `shell'-like stream of the population, possessing flattened portions in the stream arms that are reminiscent of radial fanning typically found in shells. In this context, it bears similarity to the distribution of Shell 1. We also observe `feather'-like bifurcations in different sections of the `guitar pick' shape of several streams. This is particularly evident in the profile of Stream 6 (red particles). This may be attributed to the splintering of the core in to several smaller arms which have yet to homogenise with the primary arms of the stream.

Despite the presence of several differences between the stream and shell structures, it would be difficult to classify between the two without obvious markers. The strongest indicator for each would be; a larger relative radial velocity distribution for shells, and the guitar pick shape for streams. However, as typified by Shell 2, the difference would not be obvious. Any further deviation from idealisation would further exacerbate the ambiguity.

\subsection{Action-Angle Space}

\begin{figure}[hbt!]
\centering
\includegraphics[width=\linewidth]{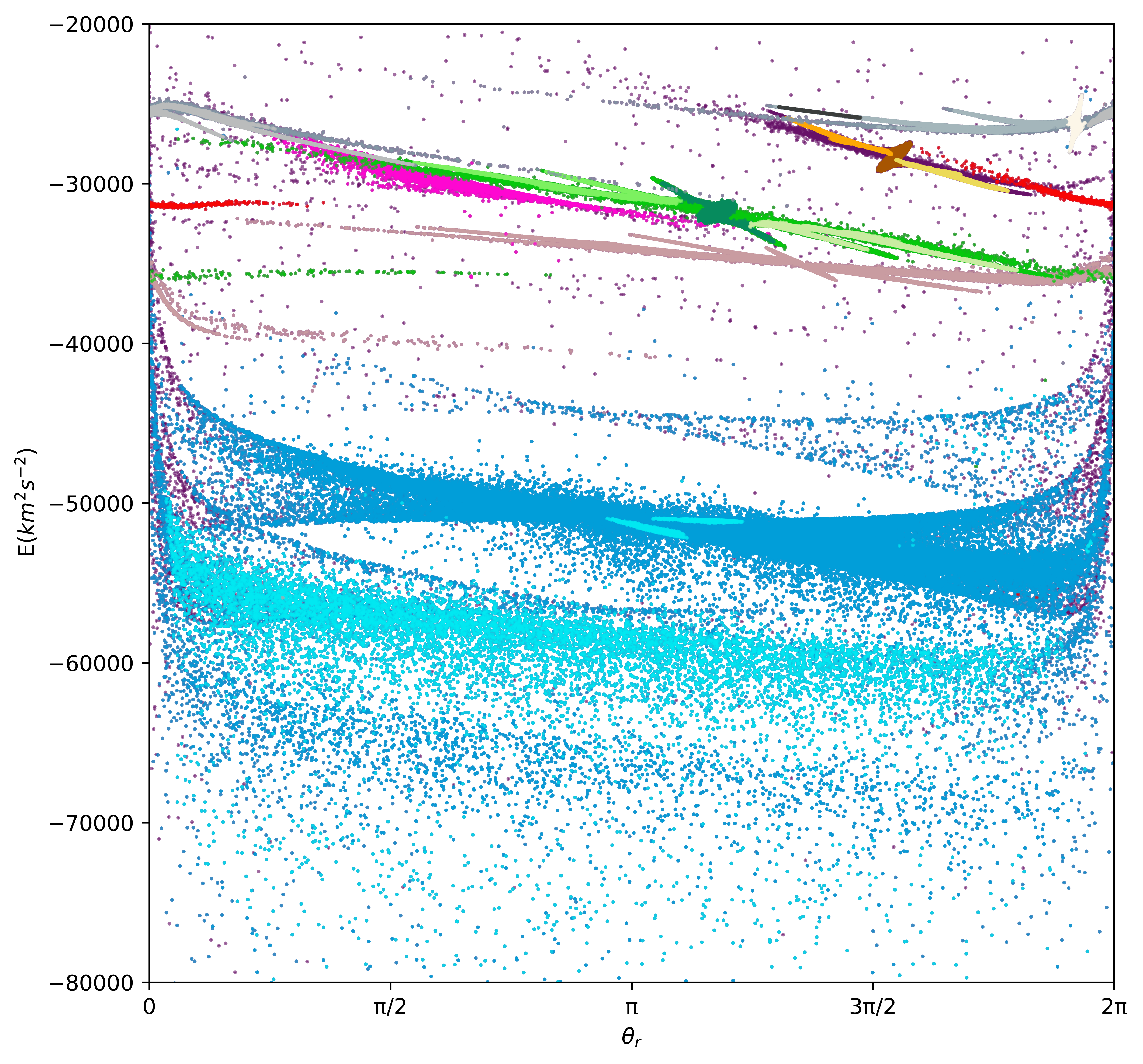}
\caption{6D clustered action-angle distribution of the t idal composite population. All cluster colours match the corresponding structures in Figure \ref{fig:four_scatter_plot}.}
\label{fig:action_angle_space}
\end{figure}
Action-angle variables provide a natural and compact coordinate system for describing orbital motion in a gravitational potential. In this formalism, the actions are integrals of motion that quantify the amplitude of oscillations along each coordinate axis, while the angles represent the phase of the motion along those axes. In a static, time-independent potential, the actions are conserved quantities, and the angles evolve linearly with time due to the separability of the Hamiltonian, we use the St\"ackel approximation to compute the action and angle variables for the dataset. The coordinate transformation was performed using \texttt{Galpy} \citep{galpy, brief_actionanlges_MW, sanders_binney_action_anlge}. From \ref{IC_table} most of the higher mass shells and streams retain tighter and more coherent features while the less massive streams produce diffused structures.   
Shell 1 displays two distinct streaks in energy, ranging from approximately –40,000 km²/s²  to –60,000 km²/s². Given the high number of particles in this system, these features likely correspond to two phases of accretion each giving rise to a sub-shell. The more populated streak around –40,000 km²/s²  corresponds to the larger, later forming shell, while the streak around 
–60,000 km²/s²  is consistent with the earlier, less massive sub-shell observed in the spatial distribution (Figure 2). This multi-modal energy structure supports a scenario in which shell 1 formed via multiple peri-centric passages, leading to successive caustics. In contrast, Shell 2 exhibits a narrower energy distribution, suggesting the entire structure was generated during a single accretion event. The limited energy spread and compact appearance in both configuration and angle space reinforce this interpretation. Despite phase mixing, shells retain localised energy overdensities in angle space \citep{6d_view_shells}.
Among the streams, Stream 1 retains a distinct head–core–tail morphology in configuration space, consistent with partial retention of its progenitor core. This spatial structure is mirrored in action space (top right corner of Figure 4), with its graded orange-yellow is visible in sequence across the energy range –20,000 to –30,000 km²/s²  between \(\sim\frac{3 \pi}{2}\) and \(2\pi\) range. Stream 2, while detected as a single structure by AstroLink, also exhibits a similar head-core-tail structure in action space (left of Stream 1’s distribution) suggesting it is further along in the disruption process. The remaining streams display uniform narrow energy spreads indicating they are well phase mixed and fully stripped, with no remnant progenitor component. Their relatively constant energies reinforce the nature of the progenitor’s orbit, with more circular orbits producing narrower energy distributions compared to shells, which exhibit a broader spread in energy \citep{Amorisco_2015, mock_streams, streams_action_potential, streams_actionanlge}.

\subsection{Clustering Classification}
\label{sec:clustering_classification}
\begin{figure}[ht!]
\centering
\includegraphics[width=\linewidth]{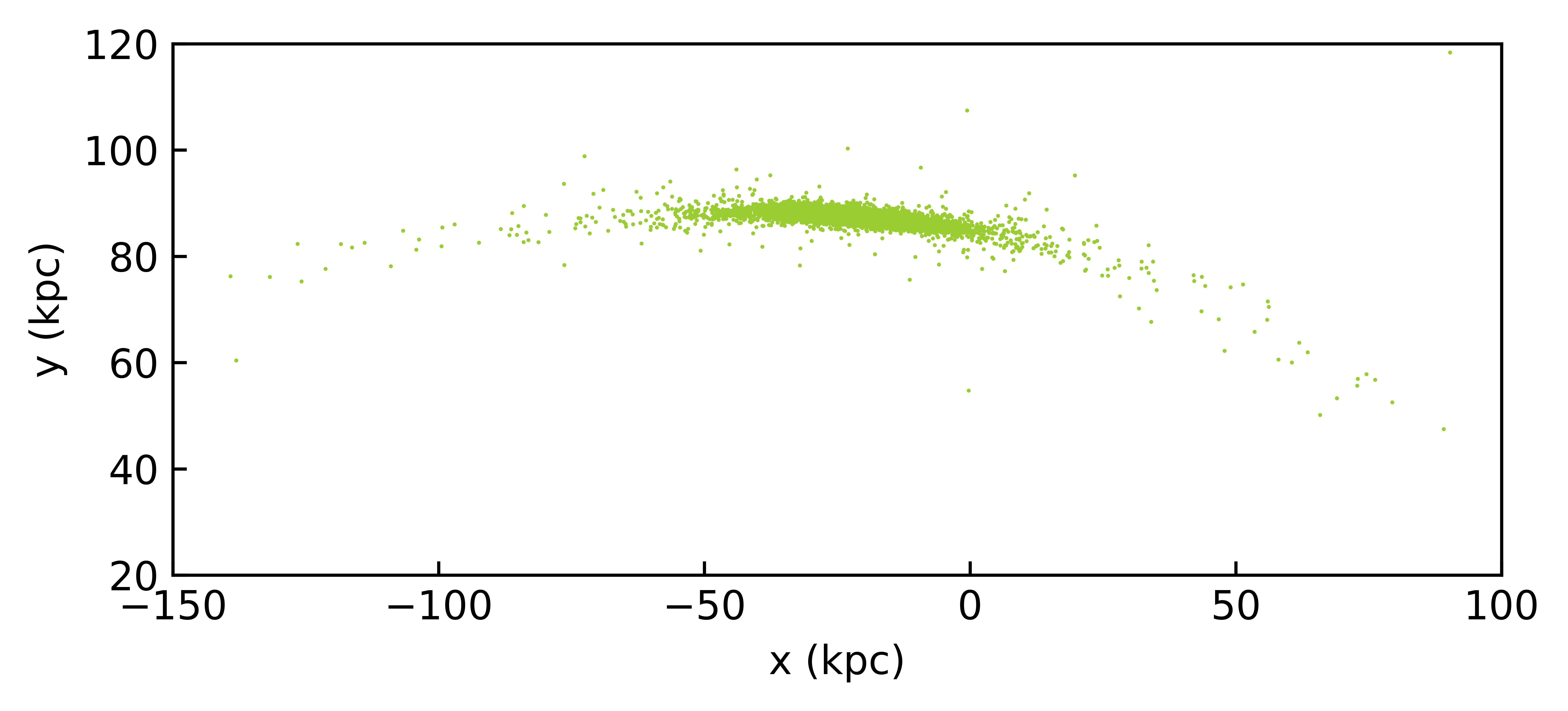}\\
\includegraphics[width=\linewidth]{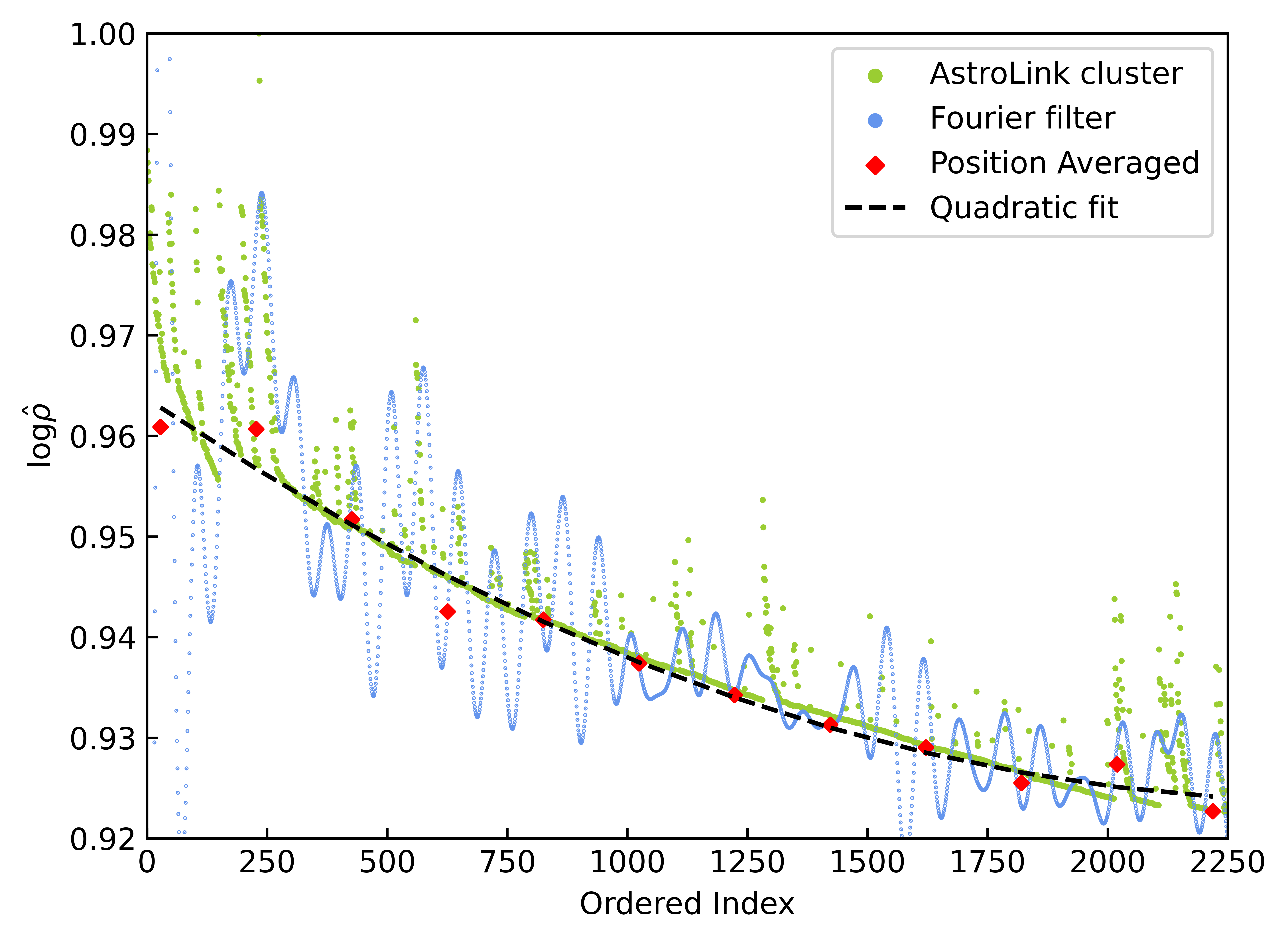}
\caption{Classification breakdown of Stream 5 as depicted in the 6D clustering and ordered density profiles in Figures \ref{fig:four_scatter_plot} and \ref{fig:odplot} respectively. The green particle spread details the initial $25\%$ of the ordered-density plot of this structure. Blue point curve characterises the filtered Fourier curve of the ordered-density plot. Red points are the position averaged values of the filtered Fourier curve. The dashed line is a depiction of the concavity test via a quadratic fit. Here we observe a positive concavity for the stream.}
\label{fig:classify_stream}
\end{figure}
\begin{figure}[htb!]
\centering
\includegraphics[width=\linewidth]{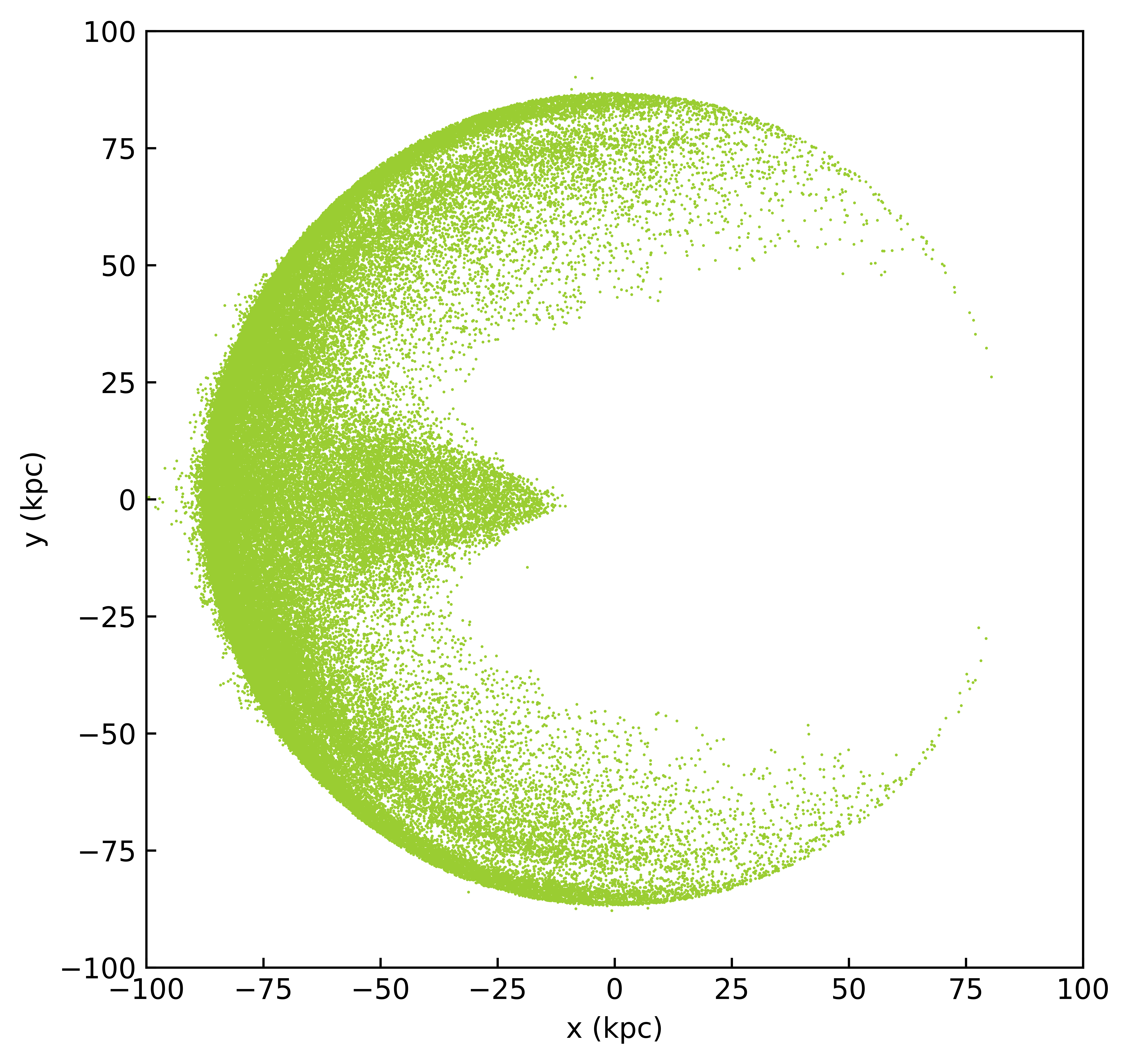}
\includegraphics[width=\linewidth]{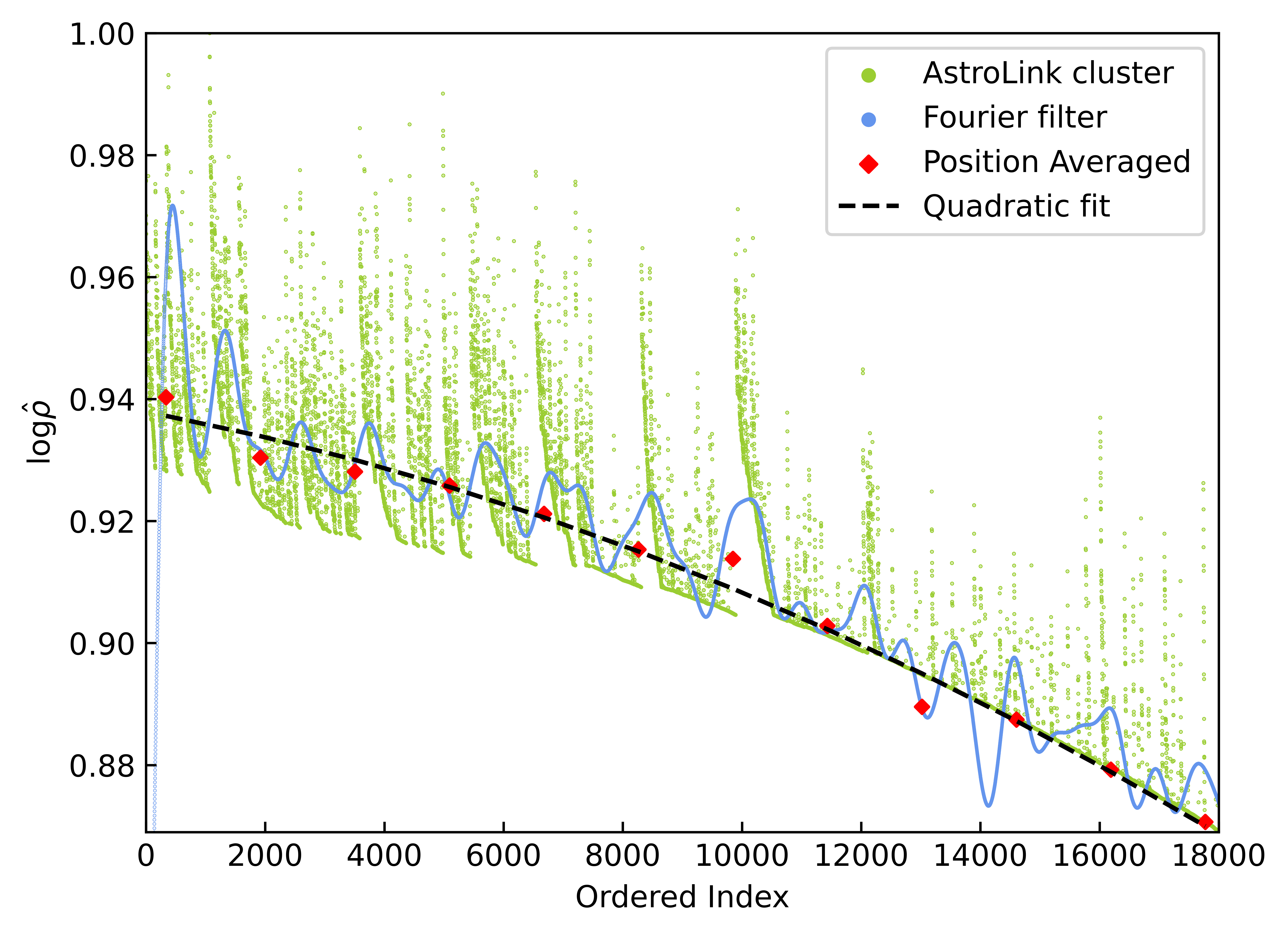}
\caption{The same method as in Figure \ref{fig:classify_stream} applied to the partial ordered density profile of Shell 1 from the 6D clustering in Figure \ref{fig:four_scatter_plot}. Here we observe a negative concavity.}
\label{fig:classify_shell}
\end{figure}

With AstroLink, we gain a novel perspective of `stream' and `shell'-like structures via the ordered-density plot. As previously stated, the clustering was conducted in 6D phase-space and thus, the ordered-density properties are influenced by both position and velocity components. The `shell' components of the tidal populations exhibit a gradual drop in density, which appears as a negative curvature in the ordered-density plot. Conversely, the stream arms exhibit a sharper drop in density, which is characterised by a positive curvature in the ordered-density plot. The contrast in these density profiles arises from the differing particle arrangements in shell and stream-like structures within phase space. Although fully developed shells exhibit a radial concentration of particles near their maximum shell radius, their greater angular particle dispersion results in a more gradual decline within the ordered-density plot. In contrast, streams have a higher particle concentration within the core remnant and arms, with little particle distribution beyond. The consequence of this is a density peak followed by a sharp density drop-off in the ordered-density plot. As such, the difference in shell and stream composition becomes apparent when viewed by their AstroLink ordered-density plot. Taking into account this difference, we are able to visually identify the `stream' and `shell'-like components solely with the ordered-density distribution of the tidal remnant population.

To take it a step further, we attempt to isolate the density curve and extract a reasonable approximation of the changing gradient. The motivation here is to demonstrate the ability to identify the quantity of `stream' and `shell'-like components in a given tidal population without the need for any form of visual or analytic inspection. Serving as an example, we design a filtered concavity test which may be applied to any specific AstroLink cluster. The first step is to take a slice of the ordered-density plot to ensure continuity without encountering density spikes from noise or secondary structures. We take an arbitrary cutoff at $4\%$ to $25\%$ index length for this purpose. Given the inherent noise present in the data, we apply a Fourier filter to the data slice profile, allowing the extraction of the underlying smoothed curve. A noise cutoff is applied to the power spectrum with respect to the maximum spectra frequency $f_\text{max}$. We take the arbitrary cutoff at $10^{-5}$ ${\rm f_{max}}$ for general application across any given ordered density slice. The filtered distribution is position averaged and applied with a quadratic fit. We observe positive concavity for `stream'-like structures and negative concavity for `shell'-like structures. Figures \ref{fig:classify_stream} and \ref{fig:classify_shell} highlight the end result of this process. From the totality of the simulations we have conducted thus far, we find no discernible correlation in the magnitude of concavity. That is, streams vary in the magnitude of concavity but are in general negative and vice versa for shells. Quantifying the difference and application into less idealistic structures remains as an avenue for future investigation.  

The application of this method was able to accurately classify all components of the composite tidal population with the exception of the stream core clusters/sub-clusters and Stream 6 which were categorised as `shell'-like structures. Despite the success, several limitations must be addressed for any future iterations involving more complex substructure presentations. Perhaps the most difficult to overcome would be particle scarcity in any particular structure. A lower particle count inherently lacks definition in the density profile, making them susceptible to misidentification. In such a case where the density plot isn't sufficiently defined, referencing radial and action space distribution and visual inspection would need to be employed. In a busier system involving background noise and larger number of substructure, we would observe a `flatter' log$ \hat{\rho}$ profile. Once again, this leads to a reduction in defined curvature for any substructure profile. In this case, a different measure of curvature may prove successful. Instead of a Fourier filter, a rolling minimum may be utilised. Essentially, for any identified cluster, the minimum log$ \hat{\rho}$ value is taken at increasing ordered index. Although this method may bias out a significant amount of data points, it may prove to be a better extraction of curvature.

\subsection{Substructure Clustering}

\begin{figure}[ht!]
\centering
\includegraphics[width=\linewidth]{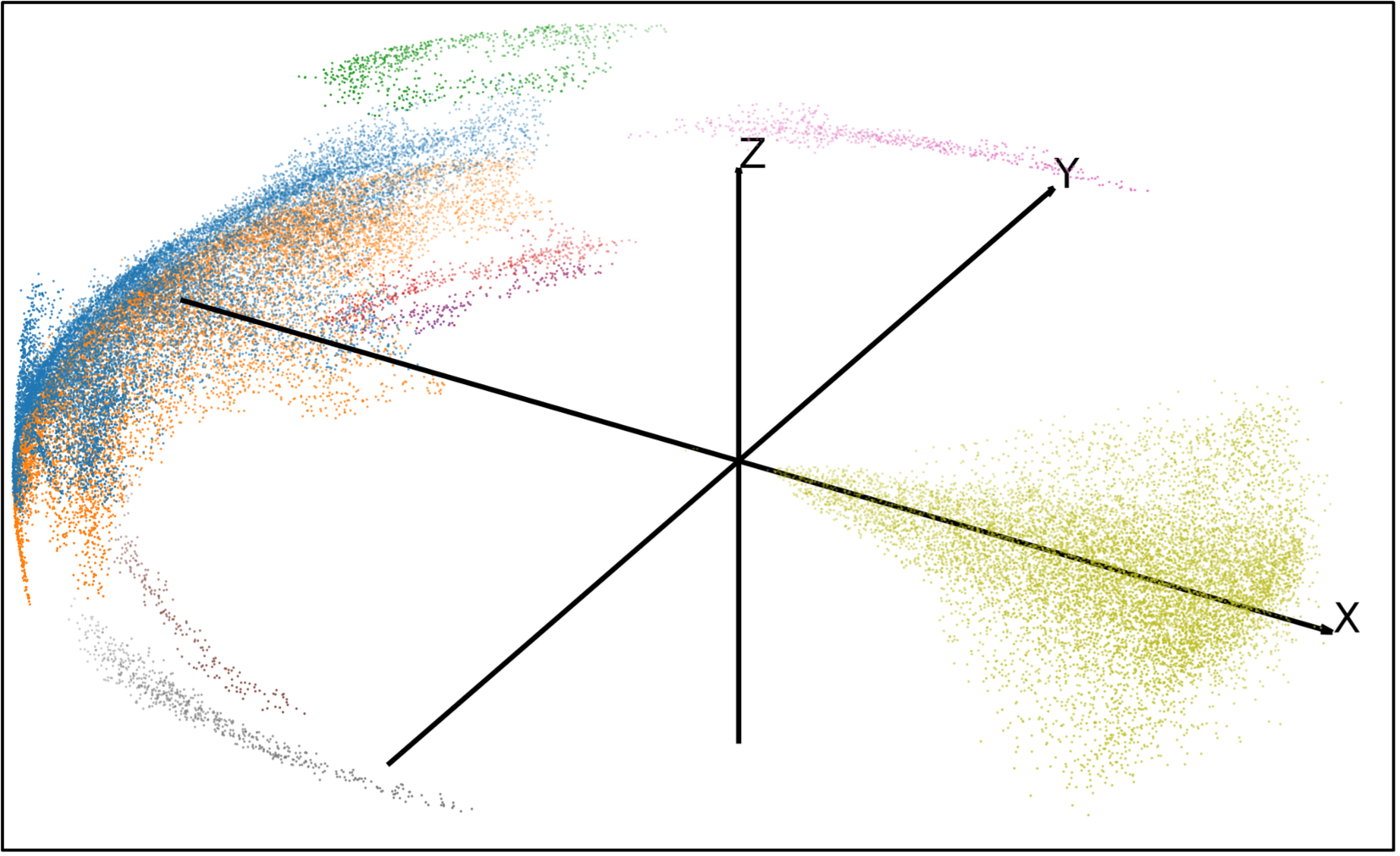}
\caption{AstroLink internal clustering of Shell 1 presented in figure \ref{fig:four_scatter_plot}. The clustering is done in 6D space. }
\label{fig:shell_substructure}
\end{figure}
\begin{figure}[ht!]
\centering
\includegraphics[width=\linewidth]{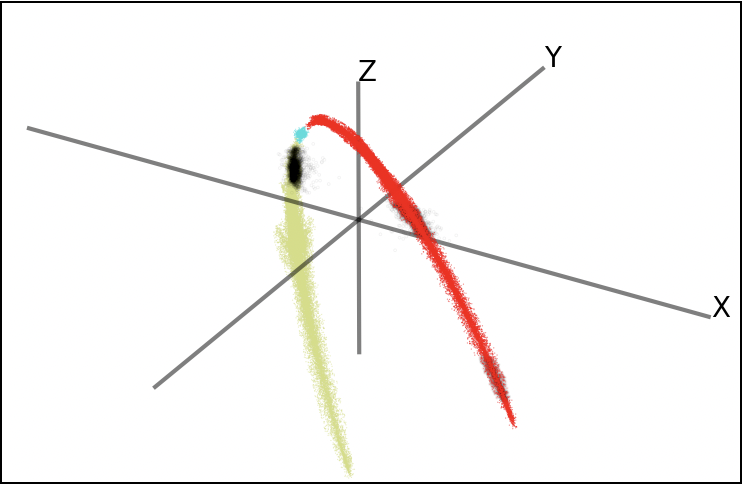}
\caption{AstroLink internal clustering of Stream 3 presented in figure \ref{fig:four_scatter_plot}. Here, the unclustered particles are displayed as the black collection of particles. The clustering is once again done in 6D space. }
\label{fig:stream_substructure}
\end{figure}

So far, we've had a clustering view from a broader perspective, identifying complete or near complete structure. With the assistance of AstroLink, we are able to delve a level deeper, investigating lower order substructure. To this end, we individually isolate Shell 1 and Stream 3 and apply AstroLink in order to explore their internal structure. The results are visualised in figures \ref{fig:shell_substructure} and \ref{fig:stream_substructure} respectively. 

Firstly, looking at our shell sample, we see clustering of the primary shell and the accompanying trailing segment. While the trailing segment is clustered as a single smooth structure, the primary shell exhibits various substructure. We are able to identify at least two major overdensities within the structure, the largest of which is the orange structure. This cluster forms the maximal arc of the shell along with the blue cluster. The separation between these two clusters is interesting and perhaps a marker of their conditions as they pass through the host centre. The portion of the progenitor below the $x-y$ plane would scatter and slingshot up along the $z$ axis as it passes the centre forming the distinct blue strip of particles. Conversely, the orange segment originates from primarily from the progenitor component above the $x-y$ plane. The asymmetry in distribution about the $z$ axis is most likely due to the influence of the galactic disk which is initialised with a one fifth flattening along the z axis. We also observe a collection of smaller clusters surrounding the two major overdensities. The green substructure stands out as it extends towards the positive z axis. Once again, its phase space distinctiveness is most likely the result of incomplete phase mixing driven by the conditions of the particle collection as they pass through the centre. The particles may be stripped away at a slightly different times or possess slightly different initial conditions which are exacerbated by the pass through the centre. It is worth appreciating the fact that we are able to cluster in 6D space. Were the shell picked up as a 2D face on projection, we would lose dimensionality and any ability to discern the separation of density regions.

The arms of the stream wrap around the central host albeit not completely. When we apply AstroLink to the isolated structure we obtain separation in key features not directly observed in the composite potential. From this perspective, we are able to see the characteristic core remnant from the progenitor along with the leading and trailing arms. Interestingly, we lose a portion of a tail in the clustering, displayed as the unclustered black particles in Figure \ref{fig:stream_substructure}. It is unclear why the portion is considered a separate to the rest of the tail. Most likely, it is due to inhomogenous phase mixing in the region where the remnant core is actively losing mass to the arm. The relatively similar length in the tails indicates that the progenitor had a symmetric encounter, i.e. the orientation of the satellite's orbital plane relative to the host's potential was aligned. 

\section{Discussion and Conclusions}
\label{sec:discussion}
Tidal remnants produced in merger events are gravitationally sensitive structures possessing a wealth of information in formation history and the nature of the ever esoteric DM. The identification and subsequent classification of these structures are an important first step in progressing their exploration. In this work, we construct a synthetic tidal remnant population evolved from Plummer spheres in an external Milky Way-like potential. The data set containing these populations is then clustered with AstroLink and the resultant structures analysed in terms of their dynamical properties. We have then also introduced a unique classification methodology which takes advantage of the clustering algorithm AstroLink. We find that the unique ordered-density profiles computed with AstroLink contain information that can be used to separate tidal `stream'-like structures from `shell'-like structures. Specifically, the ordered-density profile of `stream'-like structures has a positive curvature whilst `shell'-like structures feature a negative curvature. As such, we have demonstrated that a filtered concavity test can be used to reliably separate the two tidal structure types.
Referring to the methods outlined in \citep{StreamGen} in \ref{sec:background} and \citep{Hendel_2015}, it is evident classification typically depends on the calculation of physical parameters, which requires prior knowledge of the host potential while this approach uses the 6D information of the individual particles without needing information on the host. The absence of line-of-sight velocities (LOSV) does not significantly impact performance, as the key dynamical features are retained in 5D. Certain orientations of streams or shells relative to the observer may result in a reduced concavity signal, which could modestly impact classification strength. However, such orientations represent a limited subset of the overall configuration space. The ability to classify without the information of the host makes this method broadly applicable and compliment the physical parameter based classification.

Although we have applied our classification methodology to a synthetic sample, we find motivation for real world application in observational data studies. Most of which must rely on visual classification. \citet{johnston_2008} provided an early framework on morphological classification that grouped halo features into “mixed” (fully phase‐mixed donuts), “shells/plumes” (localized ripples), and “great‐circle” streams. ETG's contain vast range of tidal remnants that have been newly discovered or previously classed as structureless overdensity \citep{Delgado_2010, Giri_2023}. The fraction of shells and streams being 28\% and 21\%, respectively \citep{Atlas3d}.

Apart from visual classification there has been increased application of machine learning for both supervised and unsupervised classifications. Foremost are the binary classifiers that are able to distinguish between galaxies with and without tidal features \citep{Pearson_2019, Sanchez_2023}. \cite{desmons_2024} developed a self supervised learning model capable of distinguishing between galaxies with and without faint tidal structures. This model achieved strong performance using only 600 labelled examples from the HSC-SSP \citep{Aihara_2019} for training. 
Advancing this effort, \citet{CNNN_classifier} employed a convolutional neural network (CNN) trained on DECaLS \citep{Dey_2019} survey data to classify tidal features in galaxies where the structures were not overly irregular.  Training the network on streams, shells arm and diffuse structures that were visually classified prior, the classifier was also used to predict the presence and nature of tidal feature around each galaxy. Their median results for each of the structures were above 97\% of the true features. 

Over time, many observational surveys have provided data on galaxies containing these low surface brightness structures, which have been analysed to study not only the structures themselves but also the haloes of satellites and their host galaxies, shedding light on their histories. Recently \citet{NGC300} presented a detailed investigation into the accretion history of NGC 300, a galaxy with a mass comparable to that of the LMC. Their analysis reveals a complex system of tidal substructures, comprising four distinct components: two prominent stellar streams extending towards the northern and southern regions of the host galaxy, and two shell-like structures located on opposite sides of the central potential. The authors attribute all four features to a single accretion event involving one progenitor satellite, thereby demonstrating that galaxies with LMC-like masses are capable of producing rich and well-defined tidal morphologies.

With the advent of upcoming deep surveys such as LSST and the release of Gaia DR4, the potential to uncover faint and distant galactic substructures at unprecedented levels of detail is rapidly increasing. However, comprehensive analyses of individual galaxies especially those requiring visual inspection of tidal features can become prohibitively time consuming at large scales. In this context, the use of a composite gravitational potential, although idealised T , offers a valuable framework for simulating and studying the emergence of complex substructure in galaxy halos. Our application of the AstroLink demonstrates both the prevalence of `busy' subhalo environments and the algorithm’s efficiency in identifying coherent stellar substructures. Moreover, the accompanying classification scheme proves effective in distinguishing between stream-like and shell-like features, offering a scalable and automated approach to dissect the morphology of disrupted satellite debris across large datasets.  

\subsection{Future Work}
\label{sec:future_work}
With the progress made thus far, several avenues of potential future investigation open up for us. The next logical progression step is the clustering of a more complex substructure population. In this work, we have presented a curated composite potential with `idealised' tidal structures. Thus, it far detached from a realistic population that may encompass a particular galaxy halo. In particular, the absence of intermediate tidal structures which are more ambiguously defined. These may include tidal tails, umbrellas and other remnants which do not fall in any specific category. The simulation and subsequent AstroLink clustering of such a population would provide an insight into the underlying density profile for such structures. The discovery of any ordered-density features unique to these structures may provide a more robust classification system for tidal structure populations. We may achieve this through the introduction of a flattened host halo, breaking the spherical symmetry of the halo potential \citep{prolate_MW}. This introduces precession to the sub-halo orbits, influencing the evolution of their disruption. 

Further improvement involves quantifying the classification metric, we plan to incorporate the Akaike Information Criterion \citep[AIC;][]{AIC} to determine the largest fraction X of the ordered density profile within each cluster for which a quadratic fit is statistically preferred over a cubic. This approach removes the need for a user-defined threshold for X, thereby making the metric selection process more deterministic and objective. 
We plan to employ our metric to cosmological simulations containing more realistic and incomplete information of sub haloes. On the classification front, we aim to enhance the current binary scheme using machine learning techniques capable of identifying and categorising substructures in any given data sample. Given the sensitivity of tidal structures to the DM halo potential, the identification of substructure frequency would provide a novel outlook of galactic formation dynamics and by extension, dark matter.

\paragraph{Acknowledgments}
We are grateful for the feedback and comments provided by the Gravitational Astrophysics group in University of Sydney's School of Physics. We are also thankful to University of Sydney's Artemis HPC and Physics Cluster from the School of Physics for providing computational resources essential to this work. We are also grateful for the insightful comments provided by the anonymous reviewer from PASA.

\paragraph{Funding Statement}
VE is supported by the University of Sydney Postgraduate Award. SGB is supported by Australian Academy of Technological Sciences and Engineering's Elevate scholarship.  WHO's contribution to this project was made possible by funding from the Carl-Zeiss-Stiftung.

\paragraph{Competing Interests}
None

\paragraph{Data Availability Statement}
The simulation data utilised in this paper will be shared upon reasonable request to the corresponding author. The AstroLink python package is readily available at https://github.com/william-h-oliver/astrolink.

\paragraph{Ethical Standards}
The research meets all ethical guidelines, including adherence to the legal requirements of the study country.

\paragraph{Author Contributions}
Conceptualisation: S.G.B; V.E.

Methodology: V.E; S.G.B; W.H.O. Data curation: V.E. Data visualisation: V.E; S.G.B. Writing original draft: V.E; S.G.B. All authors approved the final submitted draft.
\\
\printendnotes

\bibliography{main}
\end{document}